\documentclass[pra,twocolumn,amsmath,amssymb,superscriptaddress,
showpacs]{revtex4}
\usepackage{graphics}
\usepackage{epsfig}
\usepackage{bbm}
\usepackage[latin1]{inputen c} 

\newcommand{\be}{\begin{equation}}
\newcommand{\ee}{\end{equation}}
\newcommand{\eea}{\end{eqnarray}}
\newcommand{\bea}{\begin{eqnarray}}

\newcommand{\mean}[1]{\ensuremath{\langle{#1}\rangle}}

\newcommand{\eins}{\openone}

\newcommand{\WW}{\ensuremath{\mathcal{W}}}

\newcommand{\ketbra}[1]{\ensuremath{| #1 \rangle \langle #1 |}}

\newcommand{\ket}[1]{\ensuremath{|#1\rangle}}

\newcommand{\bra}[1]{\ensuremath{\langle#1|}}

\newcommand{\kommentar}[1]{}

\renewcommand{\vr}{\ensuremath{\varrho}}

\begin{document}
\title{Toolbox for entanglement detection and fidelity estimation}
\date{\today}
\begin{abstract}
The determination of the state fidelity and the detection 
of entanglement are fundamental problems in quantum information 
experiments. We investigate how these goals can be achieved with
a minimal effort. We show that the fidelity of GHZ and W states 
can be determined with an effort increasing only linearly with 
the number of qubits. We also present simple and robust methods 
for other states, such as cluster states and states in 
decoherence-free subspaces.
\end{abstract}

\author{Otfried G\"uhne}
\affiliation{Institut f\"ur
Quantenoptik und Quanteninformation, \"Osterreichische Akademie der
Wissenschaften, 6020 Innsbruck, Austria}
\author{Chao-Yang Lu}
\affiliation{Hefei National Laboratory for Physical Sciences at
Microscale and Department of Modern Physics, University of Science
and Technology of China, Hefei, Anhui 230026, People's Republic of China.}
\author{Wei-Bo Gao}
\affiliation{Hefei National Laboratory for Physical Sciences at
Microscale and Department of Modern Physics, University of Science
and Technology of China, Hefei, Anhui 230026, People's Republic of China.}
\author{Jian-Wei Pan }
\affiliation{Hefei National Laboratory for Physical Sciences at
Microscale and Department of Modern Physics, University of Science
and Technology of China, Hefei, Anhui 230026, People's Republic of China.}
\affiliation{Physikalisches Institut, Universität Heidelberg, Philosophenweg 12,
D-69120 Heidelberg, Germany}

\maketitle


Due to recent advances in quantum control many experiments aim at the
generation and manipulation of multipartite quantum states
\cite{allexp, mohamed, lu, hartmut}. In such an experiment, one typically
aims at the creation of some pure entangled multi-qubit state $\ket{\psi},$
which can further be used for some quantum information processing task.
Due to the unavoidable noise, however, the produced state will be some
mixed state $\vr_{\rm exp},$ which may significantly differ from
the desired state $\ket{\psi}.$

First, one may be
interested in the {\it fidelity} of the produced state, that is,
to what extent the desired state was prepared. This quantity is
given by 
\be 
F_{\psi} = \bra{\psi} \vr_{\rm exp}\ket{\psi} 
\ee and
should ideally equal to one. In practice,
however, it is lower, nowadays experiments with four or more
qubits achieve typical fidelities between $0.5$ and $0.9.$

Second, one may ask whether the prepared state was
indeed {\it genuine multipartite entangled.} This means that the $N$-qubit
state $\vr_{\rm exp}$ shows entanglement effects which cannot be produced by
$N-1$ qubits and all parties must have participated in the creation
of the state $\vr_{\rm exp}$ \cite{gmpe}. To verify this
criterion of success, entanglement witnesses can be used
\cite{mohamed, wittheo, oldpra, toth, toth2, ijtp, chenchen, japanese}.
These are observables which have by construction a positive expectation
value on all unentangled (biseparable) states, a measured negative expectation
value signals the presence of entanglement.

A generic witness for a
pure state $\ket{\psi}$  is given
by
\be
\WW = \alpha \eins - \ketbra{\psi}
\label{genwit}
\ee
where $\alpha$ is the maximal fidelity of $\ket{\psi}$ for biseparable states, 
a quantity
which can be directly computed \cite{mohamed}. By construction, this observable
is positive on the biseparable states. Furthermore, from the expectation value
$Tr(\vr_{\rm exp}\WW)$ we can determine the fidelity as
$F_\psi = \alpha -Tr(\vr_{\rm exp}\WW),$ highlighting the close connection between
fidelity determination and entanglement detection \cite{berkeland}.

For the implementation, two questions are relevant. First, since quantum state 
tomography requires an exponentially increasing effort, the state $\vr_{\rm exp}$ 
is often not completely known and one has to
ask how many measurements are required for the evaluation of
$Tr(\vr_{\rm exp}\WW).$ Typically, only local measurements are possible, 
and an observable like 
\be
\mathcal{M} = \sigma_k \otimes  \sigma_k ...\otimes  \sigma_k
\ee
is called a local measurement setting \cite{oldpra, ijtp, mohamed}. Note that
measuring all the coincidence probabilities of the $2^N$ possible outcomes
of $\mathcal{M}$ gives also information about observables like
$\eins \otimes  \sigma_k ...\otimes  \sigma_k,$ etc.
A second question concerns the robustness to noise of the witness $\WW$.
As a simple model, one may consider the target state $\ket{\psi}$ mixed with
white noise, $\vr(p)=p\ketbra{\psi}+ (1-p) \eins/2^N$ and ask, how large
$p$ has to be, in order that $\vr(p)$ is detected by $\WW.$

It has been shown in  Ref.~\cite{toth} that for important classes of states,
namely GHZ states and cluster states, one can modify the witness in
Eq.~(\ref{genwit}) such that it requires only two measurement settings
and detects noisy states for $p>0.66$ (GHZ states), and this
{\it independently} of the number of qubits. 
This shows that entanglement detection and fidelity estimation get not necessarily
more difficult when increasing the number of qubits.

In this situation, three question are of interest. First,
the question arises how a witness as in Eq.~(\ref{genwit})
can be implemented by local measurements, and how many
measurements are needed. Second, the exact payoff between
the robustness to noise of a witness and the number of required
measurement settings is of interest. Finally, the question arises
whether there are constructions of witnesses beyond the
projector-based witness of Eq.~(\ref{genwit}).

In this paper we address all three questions. We first show how
the fidelity of GHZ and W states can be determined with $N+1$
(resp. $2N-1$) local measurements. Then, extending the results of
Ref.~\cite{toth} we derive witnesses for cluster states which are
robust against noise and still require a small effort. Finally,
we present a witness construction for the four-qubit singlet state,
which is simpler to measure than the construction in  Eq.~(\ref{genwit}),
and, surprisingly, more robust against noise.

Let us start by discussing the local decomposition of the
projector based witness for the case of GHZ states, 
\be 
\ket{G_N}
= \frac{1}{\sqrt{2}} (\ket{0}^{\otimes N} +\ket{1}^{\otimes N} ).
\ee 
A witness for this state is given by $\WW = \eins/2 -
\ketbra{G_N}$ \cite{mohamed}, with the term $\ketbra{G_N}$ being
decomposed as 
$
2\ketbra{G_N} = \ketbra{0}^{\otimes N}
+\ketbra{1}^{\otimes N} +\ket{0}\bra{1}^{\otimes N}
+\ket{1}\bra{0}^{\otimes N}.
$
The terms
$\ketbra{0}^{\otimes N}$ and $\ketbra{1}^{\otimes N}$ can be
directly measured with the setting $(\sigma_z)^{\otimes N}$ since
they correspond to eigenvectors of it. Using the fact that $\ket{0}\bra{1}=
(\sigma_x + i \sigma_y)/2$ and $\ket{1}\bra{0}=(\sigma_x - i
\sigma_y)/2$ we can write the remaining term
$\mathcal{X}=\ket{0}\bra{1}^{\otimes N}+\ket{1}\bra{0}^{\otimes
N}$ as \be \mathcal{X} = \frac{1}{2^{N-1}} \sum_{ k \;\;{\rm
even}} (-1)^{k/2}\sum_\pi \bigotimes_{i=1}^k \sigma_y
\bigotimes_{i=k+1}^N \sigma_x, \label{xdef} \ee
i.e. as an alternating sum of all products of the
Pauli matrices $\sigma_x$ and $\sigma_y$ with an even number of
$\sigma_y.$
Here, $\sum_\pi$ denotes
the sum over all permutations of the qubits, which yield different
expressions.

Now we choose the $N$ measurement settings
\be
\mathcal{M}_k = \big[\cos(\frac{k\pi}{N})\sigma_x
+ \sin(\frac{k\pi}{N})\sigma_y \big]^{\otimes N},
\;\;\; k = 1,...,N;
\ee
which are just measurements in the $x$-$y$-plane of the Bloch sphere
with different angles,  and obtain
$
\sum_{k=1}^N (-1)^k \mathcal{M}_k  = N \mathcal{X}
$
as can be checked by direct calculation \cite{sinus}.

{\bf Observation 1.} {\it For the determination of the fidelity of the $N$-qubit GHZ state and
evaluation on the witness $\WW = \eins/2 - \ketbra{G_N}$ the $N+1$ measurement
settings $(\sigma_z)^{\otimes N}$ and $\mathcal{M}_k$ are sufficient.}

The required number of measurements can also be determined as follows: 
$\mathcal{X}$ in Eq.~(\ref{xdef}) contains in each term only an even
number of $\sigma_y$ and $\mathcal{X}$ is symmetric under exchange 
of the qubits. The family of operators of this type is described by
$\lfloor N/2 \rfloor + 1$ parameters \cite{floor}. If we choose an 
arbitrary angle $\alpha_k \in (0, \pi/2)$  and define
$\mathcal{M}^\pm_k = [\cos(\alpha_k)\sigma_x \pm \sin(\alpha_k)\sigma_y]^{\otimes N},$
then $\mathfrak{M}_k =\mathcal{M}^+_k+ \mathcal{M}^-_k$ corresponds
to two local measurements and also contains  in each term only an 
even number of $\sigma_y$ and is symmetric under exchange of the 
qubits. Therefore, by choosing $\lfloor N/2 \rfloor + 1$ different $\alpha_k$ 
we can express {\it any} observable of the former type as a linear combination 
of the $\mathfrak{M}_k.$ Noting that we may also choose $\alpha_k =0$ 
corresponding to $(\sigma_x)^{\otimes N}$ (and, for $N$ even, also $\alpha_k =\pi/2$
corresponding to $(\sigma_y)^{\otimes N}$), we see that {\it any} observable 
which contains in each term (in the Pauli matrix representation) only an 
even number of $\sigma_y$  (and  $\sigma_z$ elsewhere) and is symmetric 
under exchange of qubits can be measured by $N$ local measurement settings.
This argumentation also shows that the measurement directions are not unique.

In order to see how this helps for other states besides GHZ
states, let us consider W states 
\cite{duerw},
defined by \be \ket{W_N} = \frac{1}{\sqrt{N}}
\sum_\pi \ket{0}^{\otimes (N-1)} \ket{1} \ee where $\sum_\pi$
denotes again the symmetrized version. 

When decomposing $\ketbra{W_N}$ into local measurements, 
the diagonal
terms can again directly be measured with the
$(\sigma_z)^{\otimes N}$ setting. The off-diagonal terms, can,
using the fact that
$\ket{01}\bra{10}+\ket{10}\bra{01}= (\sigma_x \sigma_x + \sigma_y \sigma_y)/2$
and $\ketbra{0}=(\eins+\sigma_z)/2$, be  written as
\be
\mathcal{Y}=
\frac{2^{1-N}}{N}
\Big(
\sum_\pi \bigotimes_{i=1}^2 \sigma_x  \bigotimes_{i=3}^N (\eins+\sigma_z)
+
\sum_\pi \bigotimes_{i=1}^2 \sigma_y  \bigotimes_{i=3}^N (\eins+\sigma_z)
\Big).
\ee
Each of the two terms in this sum contains only an even number of $\sigma_x$
(resp.~$\sigma_y$) and a different observable $(\eins+\sigma_z)$ elsewhere.
Hence, as discussed before, we can measure each term with $N$ measurements
settings of the type $\mathcal{M}_k^\pm = [(\cos{\alpha_k}\sigma_{x/y}
\pm \sin{\alpha_k}(\eins+\sigma_z)]^{\otimes N},$ which are effectively measurements
in the $x$-$z$-plane (resp.~$y$-$z$-plane) of the Bloch sphere. Using the fact
that one of these settings can be chosen to be $(\eins+\sigma_z)^N,$ which
is effectively a measurement of $(\sigma_z)^N,$ we can summarize:

{\bf Observation 2.}
{\it The fidelity of an $N$-qubit W state can be determined by $2N-1$ local measurements.}

Consequently, also witnesses of the type $\WW = \alpha \eins - \ketbra{W_N}$
(and also the more general witnesses in Ref.~\cite{hartmut}) can be measured
with $2N-1$ local measurements. Note that Observation 2 has general
consequences in view of the results of Ref.~\cite{chenchen}. There,
it was shown that any pure multipartite entangled state can be brought
by local operations close to the W state. Consequently, a witness of the type
$\WW = \alpha \eins - \ketbra{W_N}$, can, after reversing the appropriate
local operations, detect any entangled pure state. Then, it is proved
that the W state can be measured with $N^2-N+1$ settings, implying that
any pure entangled state can be detected by $N^2-N+1$ measurements
(although the robustness to noise may be small). Observation 2
shows that only $2N-1$ measurements are already sufficient.

The optimal decomposition for the three-qubit W state with five measurements 
was already given in Ref.~\cite{ijtp}. For the four-qubit case, we obtain the 
decomposition
\begin{align}
&\ket{W_4}\bra{W_4}=\frac{1}{64}
\big(
-2 \sum_\pi \eins \eins \eins \sigma_z-4 \sum_\pi \eins \eins \sigma_z \sigma_z
\nonumber \\
&-6 \sum_\pi \eins \sigma_z \sigma_z \sigma_z - 8 (\sigma_z)^{\otimes 4}
- 2 (\sigma_x)^{\otimes 4}- 2 (\sigma_y)^{\otimes 4}
\nonumber
\\
&+\sum_{\alpha = x,y}\big[(\eins + \sigma_z + \sigma_\alpha)^{\otimes 4}
+(\eins + \sigma_z - \sigma_\alpha)^{\otimes 4}
\big]
\big)
\end{align}
which requires the seven measurements of $(\sigma_x)^{\otimes 4}$,
$(\sigma_y)^{\otimes 4}$, $(\sigma_z)^{\otimes 4}$ and
$(\sigma_z \pm \sigma_\alpha)^{\otimes 4}$ with $\alpha = x,y.$

Let us discuss cases, where even with a simple decomposition 
the determination of the fidelity requires too much effort. This 
may happen in multi-photon experiments, where the measurement 
of a setting like $(\sigma_z)^{\otimes N}$ may require a data 
collection time of several hours. In this case, instead of 
measuring the witness $\WW = \alpha \eins - \ketbra{\psi}$ 
one can measure a witness like $\WW' = \alpha \eins - \ketbra{\psi} + P$ 
where $P$ is a positive operator. Since $P$ has only positive eigenvalues,  
$\mean{\WW}\geq 0$ implies that $\mean{\WW'}\geq 0$ so $\WW'$ is a valid
witness, and allows to estimate the fidelity via 
$F_\psi \geq \alpha - Tr(\WW'\vr_{\rm exp}).$ The question is how to 
choose $P$ in order require only few measurements, while still 
obtaining a good bound on the fidelity.

Now we show how to derive a sequence of witnesses which are more 
and more robust to noise, while still being simple to implement. 
The idea is to start from a known witness of the type 
$\WW' = \alpha \eins - \ketbra{\psi} + P$ and to systematically subtract 
terms from $P.$ We demonstrate this for the six-qubit cluster state, 
the techniques can be straightforwardly adapted to other graph states. 
The six-qubit cluster state is given by
\be
\ket{C_6} = \frac{1}{2}(\ket{000000}+\ket{111000}+\ket{000111}-\ket{111111})
\ee
This state is a graph state, corresponding after local rotations to the
H-shaped graph \cite{lu}. A typical witness for this state would be
$\WW = \eins/2 - \ketbra{C_6},$ however, the methods presented above 
yield a decomposition with 16 measurement settings. This witness detects 
a cluster state mixed with white noise for $p > 31/63 \approx 0.492.$

In order to write down witnesses which require less effort, note that the
cluster state can be described by its stabilizing operators \cite{lu, hein}.
These are
\begin{align}
g_1&= \eins \sigma_z \eins (\sigma_x)^{\otimes 3};
\;
g_2 =(\sigma_z)^{\otimes 2} (\eins)^{\otimes 4} 
;\;
g_3 = \eins (\sigma_z)^{\otimes 2} (\eins)^{\otimes 3};
\nonumber
\\
g_4&=(\sigma_x)^{\otimes 3}\eins \sigma_z \eins
;\;
g_5= (\eins)^{\otimes 3}(\sigma_z)^{\otimes 2}\eins
;\;
g_6=(\eins)^{\otimes 4}(\sigma_z)^{\otimes 2};
\end{align}
and $\ket{C_6}$ is the unique state fulfilling $g_i \ket{C_6} = \ket{C_6}.$

As the stabilizing operators describe the state, they can be 
used for the construction of witnesses.
Indeed, as shown in Ref.~\cite{toth}, 
a witness is given by
\be
\tilde{\WW}_1 = 3 \eins - 2 (\prod_{i=1}^3 \frac{g_i+\eins}{2})
- 2 (\prod_{i=4}^6 \frac{g_i+\eins}{2}).
\ee
where the two groups of stabilizing operators correspond to the 
two-colourability of the H-shaped graph \cite{toth2}.
This witness requires only two measurement settings, namely
$(\sigma_z)^{\otimes 3}(\sigma_x)^{\otimes 3}$ and
$(\sigma_x)^{\otimes 3}(\sigma_z)^{\otimes 3}.$ It detects
states mixed with white noise for $p>0.7142.$ In order to show
that $\tilde \WW$ is a witness one can directly calculate that
$\tilde{\WW}_1 - 2 \WW \geq 0, $ i.e., 
$\tilde{\WW}_1 = 2\WW + P$  \cite{toth}.

To improve this witness, let us look at $P = \tilde{\WW}_1 - 2 \WW \geq 0.$
It is easy to see that $P/2$ is a projector onto a 49-dimensional
subspace. The idea is to find a product basis such that
many vectors of this basis lie in this subspace. Then, 
these vectors can be measured with one setting and 
they can be subtracted from the witness.
In our case, the desired basis turns out to be the computational basis,
i.e., the eigenvectors of the measurement $(\sigma_z)^{\otimes 6}.$

Indeed, if we  define for three qubits
$\mathcal{A}=\eins-\ketbra{000}- \ketbra{111}$ the
operator $P-2\mathcal{A} \otimes \mathcal{A}$ has no
negative eigenvalues. Therefore, the observable
\be
\tilde{\WW}_2 = \tilde{\WW}_1
 - 2 \mathcal{A} \otimes \mathcal{A}.
\ee
is a valid witness, and requires three settings
[$(\sigma_z)^{\otimes 3}(\sigma_x)^{\otimes 3},$
$(\sigma_x)^{\otimes 3}(\sigma_z)^{\otimes 3}$
and $(\sigma_z)^{\otimes 6}$] for its measurement.
It detects states mixed with white noise for
$ p > 0.579,$ which is a reasonable improvement.

As a sidestep, note the interesting fact that the witness 
$\tilde{\WW}_2$ may be improved by local filters. Such 
filters are invertible 
operators $F_i$ on each qubit, applying then  $F=\bigotimes_{k=1}^N F_k$ to a state, 
$\vr \mapsto F \vr F^\dagger$, keeps the entanglement (or separability) of the state.
For the case of complete knowledge of $\vr$ it was demonstrated in Ref.~\cite{hartmut}, that one
can alternatively transform a witness as $\WW \mapsto F^\dagger \WW F$, and optimize 
over $F$, in order to obtain a negative expectation value. Interestingly, if  we 
restrict our attention to filters of the type
$F_i = \alpha_i \ketbra{0} + \beta_i \ketbra{1}$ the witness $F^\dagger \tilde{\WW}_2 F$ still 
needs the 
same three measurements as $\tilde{\WW}_2,$ showing that filtering can also be useful
for the case of incomplete information. 

In order to improve  $\tilde{\WW}_2$ further, we consider other eigenvectors 
of $P,$ namely
$
\ket{\phi_{1/2}} = \ket{v_k}\otimes(\ket{001}\pm\ket{110})/{\sqrt{2}};
\ket{\phi_{3/4}} = \ket{v_k}\otimes(\ket{010}\pm\ket{101})/{\sqrt{2}}$
and
$
\ket{\phi_{5/6}} = \ket{v_k}\otimes(\ket{100}\pm\ket{011})/{\sqrt{2}},
$
where $\ket{v_k}=\ket{111}$ ($\ket{v_k}=\ket{000}$) for $k$ odd (even)
and the upper signs hold for odd $k$.
Then we can define
$
\mathcal{P}_1 = \sum_{i=1}^6 \ketbra{\phi_i}
=
\mathcal{B} \otimes \mathcal{C} -\frac{1}{2} (\eins - \mathcal{A})\otimes \mathcal{A},
$
where $\mathcal{A}$ is the same as before, $\mathcal{B}=\ketbra{111}-\ketbra{000}$
and
\be
\mathcal{C}=
\frac{1}{2\sqrt{3}}
\Big[
\Big(
\frac{\sqrt{3}\sigma_x+\sigma_y}{2}
\Big)^{\otimes 3}
+
\Big(
\frac{\sqrt{3}\sigma_x-\sigma_y}{2}
\Big)^{\otimes 3}
\Big].
\ee
So we arrive at the witness
\be
\tilde{\WW}_3 =\tilde{\WW}_2 - 2 \mathcal{P}_1
\ee
which requires five measurement settings, namely the three settings for
$\tilde{\WW}_2$ and the two settings for measuring $\mathcal{B}\otimes \mathcal{C}$,
and  tolerates noise as long as $p > 0.543.$ We can go further
by considering six other vectors $\ket{\phi_{k}}$ for $k=7,...,12$,
which arise form the $\ket{\phi_{k}}$ by swapping the qubits 1,2,3
with the qubits 4,5,6.  For them, one can define a $\mathcal{P}_2$
similar to $\mathcal{P}_1,$ and one arrives at the witness:
\begin{align}
\tilde{\WW}_4 & =\tilde{\WW}_2 - 2 (\mathcal{P}_1  +\mathcal{P}_2)
\nonumber
\\
&=\tilde{\WW}_1- \eins \otimes \mathcal{A}
- \mathcal{A} \otimes \eins
- 2 (\mathcal{B} \otimes \mathcal{C}+ \mathcal{C} \otimes \mathcal{B})
\end{align}
Note that for measurement of $\tilde{\WW}_4$ the setting
$(\sigma_z)^{\otimes 6}$ is not required anymore, so in total only
six measurement settings are needed. The witness detects states
for $p>0.5$, so it is nearly as efficient as the original
projector witness $\WW = \eins/2 - \ketbra{C_6},$ however,
$\tilde{\WW}_4$ requires a significantly smaller effort. It is
this witness, which has also been used in a recent experiment
\cite{lu}.

There is another feature which makes the witness $\tilde{\WW}_4$ interesting from
a general point of view. Namely, it can be written as
\be
\tilde{\WW}_4 = \frac{\eins}{2} - \ketbra{C_6} +  \ketbra{\tilde{C}_6}
\ee
where $\ket{\tilde{C}_6} = (- \ket{000000} + \ket{000111}
+\ket{111000} +\ket{111111})/4$ is an orthogonal cluster
state in a different basis. It is immediately
clear that $\tilde{\WW}_4$ is a witness, which is of a similar
strength as the witness $\WW = \eins/2 - \ketbra{C_6}$ since
$\ket{{C}_6}$ and $\ket{\tilde{C}_6}$ are orthogonal. Also,
it becomes clear why the witness $\tilde{\WW}_4$ requires only
a moderate effort for its measurement: when comparing $\ketbra{C_6}$
and $\ketbra{C_6} +  \ketbra{\tilde{C}_6}$ in the standard basis, in the latter
some off-diagonal terms are canceled.  As we know from the analysis
of the GHZ and the W state, such off-diagonal terms are typically
difficult to measure.

Interestingly, other simple witnesses in the literature can also be viewed
in this way. For instance, for the two-qubit singlet state
$\ket{\psi^\pm} = (\ket{01}\pm\ket{10})/\sqrt{2}$  a witness would
be $\WW = \eins/2 - \ketbra{\psi^-}$ which can be simplified to
$\WW' = \eins/2 - \ketbra{\psi^-} + \ketbra{\psi^+} =
(\eins + \sigma_x\otimes\sigma_x + \sigma_y \otimes \sigma_y )/2,$
which  is known from Ref.~\cite{toth2}. For the four-qubit cluster state
$\ket{{C}_4} = (\ket{0000} + \ket{0011} +\ket{1100} -\ket{1111})/2$
we can consider again the orthogonal state
$\ket{\tilde{C}_4} = (- \ket{0000} + \ket{0011} +\ket{1100} +\ket{1111})/2$
and then the witness $\WW' = {\eins}/{2} - \ketbra{C_4} +  \ketbra{\tilde{C}_4}.$
This observable requires four measurements and has, in a different form, already
been derived in Ref.~\cite{japanese}.

{\bf Observation 3.} {\it For cluster states, one can derive sequences of witnesses, which require
more and more measurements for an implementation, while becoming more and more
robust to noise. Some of them can be improved by local filters, even if only few measurements 
have been done. Furthermore, a promising ansatz for a robust witness for a
state $\ket{\psi}$ which requires only a moderate effort, is to consider
$\WW= \alpha \eins - \ketbra{\psi} + \ketbra{\phi}$, where $\ket{\phi}$
is orthogonal to $\ket{\psi}.$}

Let us finally discuss an example how one can derive entanglement witnesses
beyond the projector based witnesses. 
We consider states in decoherence free subspaces. These are
pure states $\ket{\psi},$ which are invariant under a simultaneous unitary
rotation on all qubits, i.e. we have $(U)^{\otimes N}\ket{\psi} = e^{i\phi}\ket{\psi}$
(or, equivalently $ [\ketbra{\psi}, (U)^{\otimes N}] =0$). An example
is the four-qubit singlet state \cite{weinzuk}
\be
\ket{\Psi_4} = \frac{1}{\sqrt{3}}
\big(
\ket{0011} +\ket{1100} - \frac{1}{2}(\ket{01}+\ket{10})\otimes(\ket{01}+\ket{10})
\big)
\ee
A natural witness for this state would be $\WW=3/4 \eins - \ketbra{\Psi_4}.$
It detects the state for $p>0.733$ and requires the measurement of 15 local
settings \cite{mohamed}.

In order to write down a simpler witness, note that due to the unitary invariance of
$\ket{\Psi_4}$ we have $(\sigma_k)^{\otimes 4}\ket{\Psi_4} =
\ket{\Psi_4},$ for $k=x,y,z.$ That is, observables of the type
$g_k = (\sigma_k)^{\otimes 4}$ are stabilizing operators of
$\ket{\Psi_4}.$ Therefore, they may be a good tool to construct
entanglement witnesses. It should be noted, however, that
$\ket{\Psi_4}$ is not the only common eigenstate of the $g_k,$ the
product of two singlet states,
$\ket{\phi}=\ket{\psi^-}\otimes\ket{\psi^-}$ is a different one.

In order to distinguish between the $\ket{\Psi_4}$ and the product of two-singlets,
let us consider
\begin{align}
\mathcal{D}&(\sigma_k)= (\sigma_k)^{\otimes 4}
+ 3 \big[
(\sigma_k)^{\otimes 2}(\eins)^{\otimes 2}+
(\eins)^{\otimes 2}(\sigma_k)^{\otimes 2}
\big]
\nonumber
\\
&-\frac{3}{2}
\big[
\eins (\sigma_k)^{\otimes 2} \eins+
\eins \sigma_k \eins \sigma_k+
\sigma_k \eins \sigma_k \eins+
\sigma_k(\eins)^{\otimes 2}\sigma_k
\big]
\end{align}
Here, the coefficients have been chosen in such a way, that for the four-qubit singlet
state all terms have the expectation value one, e.g.
$3 \bra{\Psi_4}[(\sigma_k)^{\otimes 2}(\eins)^{\otimes 2}]\ket{\Psi_4}=1,$ etc.
Then we consider
$
\mathcal{Q} = \mathcal{D}(\sigma_x)+ \mathcal{D}(\sigma_y)+\mathcal{D}(\sigma_z)
$
For the singlet state we have $\bra{\Psi_4}\mathcal{Q}\ket{\Psi_4}=21.$ For
a biseparable state $\varrho_{\rm bs}$, direct numerical optimization gives
$Tr(\varrho_{\rm bs}\mathcal{Q}) \leq 14.35.$ Therefore,
$\WW = 14.35 \cdot\eins - \mathcal{Q}$ is a valid witness. This witness has
interesting properties: It requires only three measurement settings (namely
$(\sigma_k)^{\otimes 4}, k=x,y,z$) and detects noisy states for $p > 0.683.$
Therefore, despite of being simpler than the projector based witness, it is more
robust to noise, and we have:

{\bf Observation 4.} {\it For states in decoherence-free subspaces, entanglement witnesses can
be constructed from their stabilizing operators $g_k=(\sigma_k)^{\otimes N}.$
These witnesses are independent of the projector based witness. }

In conclusion, we have shown how to estimate the fidelity and detect
the entanglement for several important families of states. Our constructions demonstrate
that one can typically solve these problems with a small experimental effort. 
Therefore, our results will be useful for future experiments on multipartite entanglement.

We thank P. Hyllus, G. T\'oth and W. Wieczorek for discussions. 
This work has been supported by the FWF, the EU (OLAQUI, SCALA, 
QICS, Marie Curie Excellence Grant) and the Alexander von Humboldt 
Foundation. {This work has also been supported by NNSF, the CAS, 
and the NFRP of China.}


\begin{thebibliography}{99}

\bibitem{allexp}
Z. Zhao {\it et al.}, Nature (London) {\bf 430}, 54 (2004);
P.Walther {\it et al.}, {\it ibid.} {\bf 434}, 169 (2005);
D.~Leibfried {\it et al.}, {\it ibid.} {\bf 438}, 639 (2005);
N. Kiesel {\it et al.}, {Phys. Rev. Lett.} {\bf 95}, 210502 (2005);
{\it ibid.} {\bf 98}, 063604 (2007);
G. Vallone {\it et al.}, {\it ibid.} {\bf 98}, 180502 (2007);
K. Chen {\it et al.}, arXiv:0705.0174v1 [quant-ph].

\bibitem{hartmut}
H.~H\"affner {\it et al.},
Nature (London) {\bf 438}, 643 (2005).

\bibitem{mohamed}
M. Bourennane {\it et al.},
Phys. Rev. Lett. {\bf 92}, 087902 (2004).

\bibitem{lu}
C.-Y. Lu {\it et al.}, Nature Physics
{\bf 3}, 91 (2007).

\bibitem{gmpe} Technically speaking, a pure $N$-qubit state
$\ket{\psi}$ is called biseparable, iff it can be written
as $\ket{\psi}=\ket{\alpha}\otimes\ket{\beta}$ where
$\ket{\alpha}$ and  $\ket{\beta}$ are states on $M$ and $N-M$
qubits. A mixed state is biseparable, if it can be written as
a convex combination of biseparable pure states, and  a state which
 is not biseparable, is called genuine multipartite entangled.
See also Refs.~\cite{mohamed, ijtp}.

\bibitem{wittheo}
M. Horodecki, P. Horodecki, and R. Horodecki,
Phys. Lett. A {\bf 223}, 1 (1996);
B. Terhal, Phys. Lett. A {\bf 271}, 319 (2000);
M. Lewenstein {\it et al.},
Phys. Rev. A {\bf 62}, 052310 (2000);
D. Bru\ss~{\it et al.},
J. Mod. Opt. {\bf 49}, 1399 (2002);
O. G\"uhne and N. L\"utkenhaus,
Phys. Rev. Lett. {\bf 96}, 170502 (2006);
S.J. van Enk, N. L{\"u}tkenhaus and
H.J. Kimble, Phys. Rev. A {\bf 75}, 052318 (2007);
G. T\'oth, J. Opt. Soc. Am. B {\bf 24}, 275 (2007).

\bibitem{oldpra}
B.M. Terhal, Theoret. Comput. Sci. {\bf 287}, 313 (2002);
O. G\"uhne {\it et al.}, Phys. Rev. A {\bf 66}, 062305 (2002).

\bibitem{ijtp}
O. G{\"u}hne and P. Hyllus, Int. J. Theor. Phys. {\bf 42}, 1001 (2003).

\bibitem{toth}
G. T\'oth and O. G\"uhne,
Phys. Rev. Lett. {\bf 94}, 060501 (2005).

\bibitem{toth2}
G. T\'oth and O. G\"uhne,
Phys. Rev. A {\bf 72}, 022340 (2005).

\bibitem{japanese}
Y. Tokunaga {\it et  al,},
Phys. Rev. A {\bf 74}, 020301(R) (2006).



\bibitem{chenchen}
L. Chen and Y.-X. Chen, Phys. Rev. A {\bf 76}, 022330 (2007).


\bibitem{berkeland} R.D. Somma, J. Chiaverini, and D.J. Berkeland,
Phys. Rev. A {\bf 74}, 052302 (2006).

\bibitem{sinus} For this calculation, the identities
$\sum_{k=1}^N (-1)^k \cos^m(k\pi/N) = (1+(-1)^{N+m})/2$
(for $m<N$) and $\sum_{k=1}^N (-1)^k \cos^m(k\pi/N) =
N/2^{N-1}$ (for $m=N$) are of great help. See A. Prudnikov, Y. Brychkov
and O. Maridev, {\it Integrals and Series, Vol. 1},  Taylor and Francis (1998),
p. 640.

\bibitem{floor} The symbol $\lfloor x \rfloor$  denotes
the largest integer which is smaller or equal $x$.

\bibitem{duerw} W. D\"ur, G. Vidal, and I. Cirac, 
Phys. Rev. A {\bf 62}, 062314 (2000).

\bibitem{hein}
M. Hein, J. Eisert, and  H.-J. Briegel,
Phys. Rev. A {\bf 69}, 062311 (2004).

\bibitem{weinzuk}
H. Weinfurter and M. \.Zukowski, Phys. Rev. A, {\bf 64}, 010102 (2001);
A. Cabello, Phys. Rev. A {\bf 75,} 020301 (2007).

\end{thebibliography}
\end{document}